# Acoustic contributions of a sound absorbing blanket placed in a double panel structure: Absorption Versus Transmission

Olivier Doutres and Noureddine Atalla

*GAUS, Department of mechanical engineering, Université de Sherbrooke (Qc), Canada, J1K 2R1*

Olivier.Doutres@USherbrooke.ca

Running title:  Simplified double panel sound transmission loss.

**ABSTRACT**

The objective of this paper is to propose a simple tool to estimate the absorption vs. transmission loss contributions of a multilayered blanket unbounded in a double panel structure and thus guide its optimization. The normal incidence airborne sound transmission loss of the double panel structure, without structure-borne connections, is written in terms of three main contributions; (i) sound transmission loss of the panels, (ii) sound transmission loss of the blanket and (iii) sound absorption due to multiple reflections inside the cavity. The method is applied to four different blankets frequently used in automotive and aeronautic applications: a non-symmetric multilayer made of a screen in sandwich between two porous layers and three symmetric porous layers having different pore geometries. It is shown that the absorption behavior of the blanket controls the acoustic behavior of the treatment at low and medium frequencies and its transmission loss at high frequencies. Acoustic treatment having poor sound absorption behavior can affect the performance of the double panel structure.

## I. INTRODUCTION

The sound transmission performance of double panel systems containing sound absorbing blankets is of utmost importance for noise control in various applications. In this structure, some of the relevant variables are the mechanical properties of each panel, the air gap separating the two panels, the effects of the absorbing blanket, the frequency of excitation and the type of source. The objective of this paper is to propose a simple tool to estimate the various acoustical contributions of the absorbing blanket inside the air gap, with an eye to facilitate the acoustic treatment optimization. The tool is limited to normal incidence and laterally infinite systems.

The effect of an absorptive layer in the air gap has been widely studied, both theoretically and experimentally. Beranek and Work[1] and later Fahy[2,3] proposed a wave decomposition method in which the absorbing material was considered as an equivalent fluid (rigid or limp)[4]. The normal sound incidence transmission loss of the double panel structure is derived by assigning a complex acoustic wave number to the air filling the cavity. It is shown that the absorbing blanket plays a role mainly at high frequencies where the acoustic resonance effects inside the structure are minimized and the sound transmission loss increased. Compared to experiments, this model exhibits good agreement except in the frequency range where the porous frame vibration is important[5]. Moreover, Fahy[3] mentioned that this model does not allow a straightforward explication of the effect of the absorber parameters on sound transmission by means of parametric approximations as it is possible in the case of the empty cavity. In the same way, Gösele [6] proposed a simplified method to predict the sound transmission loss through a double wall, without structure-borne connections and where the air gap is filled with porous sound-absorbing material. Denoting the measured sound transmissions of the two constituent single partitions by $TL_{p1}$ and $Tl_{p2}$, the transmission loss of the double wall system is approximated by[6]

$$TL = TL_{p1} + TL_{p2} + 20\log\left(\frac{2\omega Z_0}{s'}\right),$$  (1)

with $Z_0$ the characteristic impedance of ambient air, $\omega$ the angular frequency and $s'$ the dynamic stiffness of the gap given for different frequency ranges. However, this simple model does not allow to account for the blanket contribution at the acoustic resonance frequencies of the cavity. Later, Bolton [7] extends the wave decomposition method to the random incidence transmission loss of infinite lined panel structures. The frame influence of the porous layer is taken into account using a poroelastic model based on Biot's theory [8]. Bolton shows that the transmission loss is generally improved in the high frequency range when the porous layer is not bonded on the vibrating panels, i.e. with no mechanical coupling between the panels and the blanket. However, the transmission loss performance at low frequencies could be increased at the expense of high frequencies by bonding the porous layer onto one of the panels. Allard[4] and Lauriks[9] proposed a method based on transfer matrices to account for the porous layer in infinite double-leaf partitions. This Transfer Matrix Method (called TMM in this paper) is adequate for describing layered partitions, composed of infinite plates and porous layers, with acceptable accuracy[5], the porous layers being modeled either with an equivalent fluid or a poroelastic model. However, all these studies deal with laterally infinite double-panel systems and are based on non-modal methods. Consequently, the respective conclusions may not be convenient for finite double-plate systems, especially in the low-frequency range where the modal behaviour is manifest. Note that extensions to the TMM for finite size panels are discussed in ref.[4] and are shown to capture well the low frequency behaviour of the system. To better assess the modal behaviour and mounting conditions at low frequencies, Panneton and Atalla[10] developed a three-dimensional finite element model to predict the sound transmission loss through finite multilayer systems made of elastic, acoustic, and poroelastic media. Applied to the airborne transmission loss of a double panel structure, they shown that the low frequency transmission loss is optimum when the porous layer in bonded only on one of the two panels; the porous layer providing at the same time mechanical and acoustical damping to the structure.

These different models have also been used to optimize the effect of the porous layer and increase the transmission loss performance of the double panel structure. The optimization was mainly based on the blanket configuration[7,10,11,12], i.e. thickness, position and boundary conditions, or on the properties of the porous layer such as static airflow resistivity or bulk density[3,7,11].

Even though these aforementioned models allow one to predict quite accurately the effect of a blanket on the sound transmission loss of the double panel structure, it remains unclear what acoustical properties the absorbing blanket should own to optimize the sound transmission loss of the double panel structure. Indeed, considering an unbounded blanket inside the cavity, this layer will dissipate acoustical energy by means of two main contributions: sound absorption in the air-gaps separating the blanket from the panels and sound transmission loss through the blanket. Thus, the knowledge of how much these two different properties contribute to the sound transmission loss could help one to develop an adequate absorbing blanket. Indeed, an efficient material in terms of sound transmission loss can show poor sound absorbing behaviour, and vice versa.

The objective of this paper is to propose a simple tool to estimate the sound absorption and sound transmission loss contributions of the blanket on the sound transmission loss of the double panel structure. Since this work mainly focuses on the acoustic contribution of the blanket, only the airborne transmission of an infinite double panel structure filled with an unbounded absorbing blanket and excited by plane waves is considered. The unbounded case, in real configurations, occurs when the blanket is not fully bonded onto the vibrating panel. Practically, a 1mm air-gap is set between the panel and the blanket to simulate this decoupling. A larger air gap can also be intentionally devised to optimize the transmission loss. For this purpose, a simple analytic expression of the normal incidence sound transmission loss of the double panel structure is proposed here in terms of the three main contributions: (i) sound transmission loss of the panels, (ii) sound transmission loss of the blanket and (iii) sound absorption due to multiple reflections inside the air-gaps. In the first part, this simple analytical expression and its various contributions are presented. Next, applications to four different absorbing blankets frequently used in transport applications and the weight of each contribution are investigated.

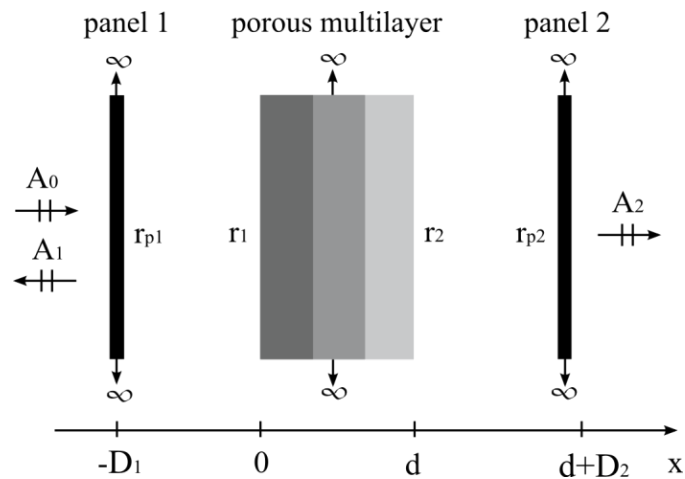

*Figure 1 : Schematic view of the double panel structure.*

## ll. THEORY

### A. General formulation

A schematic view of the structure is shown in Fig. 1. This structure is a partition consisting of two thin homogeneous panels, separated by an air-gap containing an acoustically absorbing multilayer blanket, and with no interconnections between the two panels. This configuration is typical in the automotive or aeronautic context. The air layer between the first panel and the front face of the porous multilayer has a thickness $D_1$ and is called the upstream cavity in this paper. The air layer between the rear face of the porous layer and the second panel has a thickness $D_2$ and is called the downstream cavity. The porous multilayer thickness is denoted by $d$.

A plane wave impinges on the first panel with the amplitude $A_0$, it is attenuated as it goes through the double panel structure and is finally reduced to the amplitude $A_2$ in the receiving semi-infinite air domain ($x > d + D_2$). The normal incidence transmission coefficient and the TL of the system can be expressed as, respectively

$$t = \frac{A_2}{A_0},$$ 

(2)

and

$$TL = -20\log\left(\left|\, t\,\right|\right).$$ 

(3)

According to Bruneau[13], the total sound pressure field between two planes characterized by an acoustic surface impedance can be written in terms of two plane waves, the distance separating the planes and the two reflection coefficients. Applied to the upstream cavity (*-D₁* $<x<0$) and the downstream cavity ($d<x<d+D_2$) inside the double panel structure, it is given by, respectively (the *exp(jωt)* time dependence is omitted)

$$p_u(x) = \frac{a}{1 - r_1 r_{p1} e^{-2jk_0 D_1}} \left( e^{-jk_0 x} + r_1 e^{jk_0 x} \right),$$ (4)

$$p_d(x) = \frac{b}{1 - r_2 r_{p2} e^{-2jk_0 D_2}} \left( e^{-jk_0 x} + r_{p2} e^{-2jk_0(d+D_2)} e^{jk_0 x} \right).$$ (5)

This wave decomposition method takes explicitly into account the multiple wave reflections in the upstream and downstream cavities and makes no assumption on the reflection coefficients. Here, $k_0$ is the wave number in the ambient fluid, $r_1$ the reflection coefficient seen by the x-positive propagating waves in the upstream part, $r_2$ the reflection coefficient seen by the x-negative propagating waves in the downstream part and $r_{p1}$ and $r_{p2}$ the reflection coefficients of the first and second panels respectively. A definition and a simple analytical expression of these coefficients will be presented later in the paper.

The two coefficients *a* and *b* are respectively the amplitude of the first acoustic wave transmitted by the panel 1 in the upstream cavity and the amplitude of the first acoustic wave transmitted by the porous multilayer in the downstream cavity. The amplitude coefficient *a* is thus simply the incident wave $A_0$ attenuated as it goes through the first panel by the complex sound transmission coefficient $t_{p1}$: $a=A_0\,t_{p1}$. According to Eq.(4), the total sound pressure impinging on the front face of the porous multilayer at $x=0$ is $A_0 t_{p1}/(1- r_1 r_{p1}\exp(-2jk_0 D_1))$. In the same way, this amplitude is attenuated as it goes through the porous multilayer by the complex sound transmission coefficient $t_m$ and *b* is given by $b=A_0 t_{p1} t_m/(1- r_1 r_{p1}\exp(-2jk_0 D_1))$. According to Eq. (5), the total incident sound pressure on the downstream part at $x=d+D_2$ is $[A_0 t_{p1} t_m/(1-$ $r_1r_{p1}\exp(-2jk_0D_1))(1-\ r_2r_{p2}\exp(-2jk_0D_2))]$. This sound pressure is finally attenuated as it goes through the second panel by the complex sound transmission coefficient $t_{p2}$. The normal incidence sound transmission loss of the structure is finally given by

$$TL = TL_{p1} + TL_{p2} + TL_m + TL_u + TL_d \qquad \text{dB}, \tag{6}$$

with $TL_{p1}$=-20log($|t_{p1}|$), $TL_{p2}$=-20log($|t_{p2}|$), $TL_m$=-20log($|t_m|$),

$$TL_u = -20\log\left(\frac{1}{\left|1 - r_1 r_{p1} e^{-2jk_0 D_1}\right|}\right), \tag{7}$$

and

$$TL_d = -20\log\left(\frac{1}{\left|1 - r_2 r_{p2} e^{-2jk_0 D_2}\right|}\right). \tag{8}$$

Eq. (6) shows that the normal incidence sound transmission loss of the multilayer can be decomposed into three main parts. $TL_{p1}$ and $TL_{p2}$ account for the sound transmission loss of the first and second panel, $TL_m$, accounts for the sound transmission loss of the porous layer and finally, $TL_u$ and $TL_d$, account for the multiple wave reflections in the upstream and downstream cavities inside the double panel structure, respectively. The weight of each contribution to the global sound transmission loss will be illustrated later in this paper for a multilayer sound absorbing blanket and three different monolayer porous materials.

**B. Calculation of the intermediate transmission loss**

*1. Panels transmission loss contribution*

The normal incidence transmission loss of each panel, $TL_{pj}$ (j=1,2), is simply derived from its surface density $m_{sj}$ as

$$TL_{pj} = -20\log\left(\left|\frac{2Z_0}{2Z_0 + j\omega m_{sj}}\right|\right),$$
(9)

with $Z_0$ the characteristic impedance of ambient air and $\omega$ the angular frequency.

*2. Blanket transmission loss contribution*

In this paper, the porous layers constituting the blanket are assumed to be acoustically rigid or limp[4,14,15] and are described from their characteristic wave number $k$ and characteristic impedance $Z_c$. These intrinsic properties are determined from the equivalent fluid model of Johnson et al.[16] and Allard and Champoux[17] which involves the measurement of the following non-acoustic properties[18]: static airflow resistivity $\sigma$, porosity $\phi$, tortuosity $\alpha_\infty$, viscous characteristic length $\Lambda$, and thermal characteristic length $\Lambda$'. If the blanket is a multilayer composed of a stack of different porous and non-porous materials (screen, heavy layer ...), the transmission loss $TL_m$, is determined from the transfer matrix method TMM[4]. Note that the assumption of rigid/limp behaviour doesn't limit the scope of Eqs. (6) to (8) since the TMM can also be used to account for poroelastic behaviour (estimation of $TL_m$, $r_1$ and $r_2$) and other type of panels (estimation of $TL_{pj}$).

In the simple case of a rigid or limp porous monolayer that is both symmetric and homogeneous, the sound transmission coefficient can be determined from the transfer matrix coefficients $T_{ij}$ as

$$TL_m = -20\log\left(\left|\frac{2e^{jk_0 d}}{T_{11} + \dfrac{T_{12}}{Z_0} + Z_0 T_{21} + T_{22}}\right|\right),$$ (10)

with

$$\begin{bmatrix} T_{11} & T_{12} \\ T_{21} & T_{22} \end{bmatrix} = \begin{bmatrix} \cos(kd) & jZ_c \sin(kd) \\ j\sin(kd)/Z_c & \cos(kd) \end{bmatrix}.$$ (11)

It is worth mentioning that this transmission loss can also be measured in an impedance tube, the porous element being symmetric[18,19] or non-symmetric[20].

### 3. Upstream and Downstream absorption contributions

According to Eqs. (7) and (8), the sound transmission loss coefficients, $TL_u$ and $TL_d$, which account for the absorption mechanisms in the upstream and downstream cavities can be determined from the reflection coefficients ($r_1$, $r_{p1}$) and ($r_2$, $r_{p2}$), respectively.

The coefficients $r_{pj}$ (j=1,2) are the reflection coefficients of a air-panel-air interface and are simply derived considering the continuity of the normal particle velocity at the front and rear face of the panel and by applying the Newton law to the panel. Finally, these coefficients are given by

$$r_{pj} = \frac{j\omega m_{sj}}{2Z_0 - j\omega m_{sj}} \ .$$

(12)

To define the two reflection coefficients $r_1$ and $r_2$, let us go back to the proposed expression of the sound transmission loss of the double panel structure. According to the figure 2(a), the sound transmission loss of the structure $TL$ can be derived from $TL=TL_{p1}+TL_u+TL_S$; $TL_S$ being the sound transmission loss of the sub-system $S$ composed of the porous multilayer, the downstream cavity and the second panel (see Fig. 2(b)).

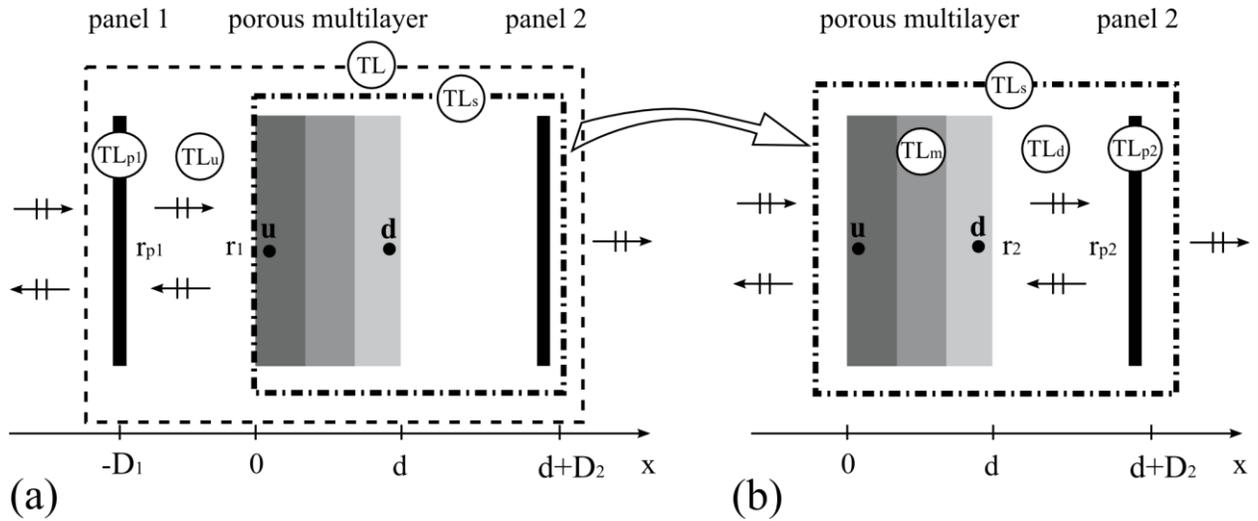

*Figure 2 : Schematic view of the double panel structure split in two parts: definition of $r_1$ and $r_2$.*

In the upstream cavity shown in Fig. 2(a), the reflection coefficient at the left hand side of the cavity is the reflection coefficient of panel 1 $r_{p1}$ (see Eq. (12)) and the one of the right hand side is the reflection coefficient of the sub-system "multilayer/downstream cavity/panel 2", $r_1$. Note that, according to the Fig. 2(a), the side **u** of the porous multilayer is facing the incident sound wave taken place in the upstream cavity; what is an important consideration for the calculation of $r_1$. This reflection coefficient can be determined using the TMM method[4].

However, it could be simplified by considering the second panel rigid and immobile. In this case $r_1$ is called $r_1{}^w$, and for a symmetric and homogeneous porous monolayer, it is given by

$$r_1{}^w = \frac{T_{11}T_{11}^a + T_{12}T_{21}^a - Z_0 T_{21}T_{11}^a - Z_0 T_{22}T_{21}^a}{T_{11}T_{11}^a + T_{12}T_{21}^a + Z_0 T_{21}T_{11}^a + Z_0 T_{22}T_{21}^a}, \tag{13}$$

with $T_{ij}$ the transfer matrix coefficients of the porous layer and $T_{ij}^a$ the transfer matrix coefficients of the downstream cavity of thickness $D_2$. Note that the coefficients $T_{ij}^a$ can be obtained from Eq. (11) by replacing $Z_c$ by $Z_0$ and $k$ by $k_0$. The simplification $r_1=r_1{}^w$ will mainly lead to a discrepancy in the estimation of the mechanical behaviour of the double panel structure in the low frequency range around the "mass/spring/mass" resonance frequency. This consideration will be detailed in the next section. It is worth mentioning that the reflection coefficient $r_1{}^w$ can also be measured in an impedance tube according to the standard ISO-10534-2[21]; the side **u** of the porous multilayer facing the incident sound wave.

Considering now the sub-system $S$ shown in Fig. 2(b), the sound transmission loss $TL_S$ is given by $TL_S=TL_m+TL_d+TL_{p2}$. Here, the downstream cavity is separated from the acoustic source ($x<0$) by the porous multilayer. Thus, the reflection coefficient at the left hand side of the cavity is the reflection coefficient of the porous multilayer backed by an infinite air layer $r_2$ and the one of the right hand side is the reflection coefficient of the second panel $r_{p2}$ (see Eq. (12)). Note that, according to Fig. 2(b), the side **d** of the porous multilayer is now facing the incident sound wave inside the downstream cavity; which is an important consideration for the calculation of $r_2$. This reflection coefficient can be determined from the TMM method[4]. However, in the case of a symmetric and homogeneous porous monolayer, it is simply given by[19]

$$r_2 = \frac{T_{11} + T_{12}/Z_0 - Z_0 T_{21} - T_{22}}{T_{11} + T_{12}/Z_0 + Z_0 T_{21} + T_{22}}.$$  (14)

The reflection coefficient $r_2$ can also be measured in an impedance tube according to the standard ISO-10534-2[21], making use this time of an anechoic termination; the side **d** of the porous multilayer faces the incident sound wave.

*Table I. Properties of the material samples*

| Material properties | Material A | Material B | Material C | Screen |
|---|---|---|---|---|
| Porosity $\varphi$ | 0.99 | 0.99 | 0.99 | - |
| Density $\rho$ (kg/m$^3$) | 7.5 | 6.1 | 5.5 | 125 |
| Static airflow resistivity $\sigma$ (Ns/m$^4$) | 7 300 | 25 000 | 14 000 | 50 000 |
| Tortuosity $\alpha_\infty$ | 1 | 2.8 | 1 | - |
| Viscous length $\Lambda$ ($\mu$m) | 88 | 100 | 70 | - |
| Thermal length $\Lambda$'($\mu$m) | 160 | 300 | 107 | - |

## III. SIMULATIONS

### A. Transmission loss contributions for a multilayer blanket

First, the validity of the proposed formulation is checked by comparison of a full TMM solution in the case of a double panel configuration close to the aeronautical context. Here a 50.5mm-thick multilayer blanket is paced between two aluminum panels with a 25mm-thick upstream cavity and a 30mm-thick downstream cavity. The two aluminum panels have the same density of 2742 kg/m$^3$ but panel 1 is 1mm-thick and panel 2 is 2mm-thick ($m_{s1}$= 2.742 kg/m$^2$, $m_{s2}$=5.484 kg/m$^2$). The multilayer blanket is made of a 0.5mm-thick resistive screen in sandwich between a 25mm-thick plastic foam (material A) and a 25mm-thick fibrous material (material C). Properties of the materials are given in table 1. The equivalent fluid limp model[14,15] is used here to describe the acoustic behaviour of the porous materials.

Fig. 3(a) presents the normal incidence sound transmission loss of the double panel structure derived from Eq. (6) considering $r_1$ or $r_1^w$ in Eq. (7), the one obtained from the TMM model and taken here as reference, the contributions of the two panels ($TL_{p1}+TL_{p2}$) derived from Eq. (9) and the normal incidence sound transmission loss of the double panel with an empty cavity. Fig. 3(b) presents the transmission loss contribution of the blanket ($TL_m$) and the absorption contributions ($TL_u$, $TL_d$) derived from Eqs. (7) and (8) when the blanket is present or not in the air-gap. Finally, Fig. 4 presents the reflection coefficients $r_1$, $r_1^w$, $r_2$ and $r_{pj}$ (j=1,2) required for the $TL_u$ and $TL_d$ calculations. The associated absorption coefficients are also presented (row 1) and are derived from the reflection coefficients $r$ as $\alpha$=1-$|r|^2$. Since the sound absorbing blanket is a multilayer in this configuration, the transmission loss of the blanket $TL_m$ and the reflection coefficients $r_1$, $r_1{}^w$, and $r_2$ are derived from the TMM model. $r_{pj}$ is given by Eq.(12) knowing the surface density of the plates.

Fig. 3(a) shows that the proposed expression of *TL* (Eq. (6)) associated to the reflection coefficient $r_1$ gives the same result compared to the reference TMM model; this corroborates the validity of the simple expression of *TL* and of the calculation of the various reflection coefficients. As mentioned previously, the use of the simplified coefficient $r_1{}^w$ in Eq. (7) leads to a discrepancy in the estimation of the mechanical behaviour of the double panel structure in the low frequency range around the "mass/spring/mass" resonance frequency (up to 300 Hz). However, this simplification has no effect in the medium and high frequency bands. Indeed, Fig. 4 shows that the absorption behaviour of the sub-system "multilayer/downstream cavity/panel 2", associated to $r_1$, is close to the absorption of the system for which the second panel would be replaced by a rigid and immobile surface (see column 1). The main difference is the high absorption coefficient at low frequencies due to the dynamic behaviour of the second panel; this absorption behaviour being characteristic for the two plates as shown in column 3 (absorption coefficient related to the reflection coefficients $r_{pj}$).

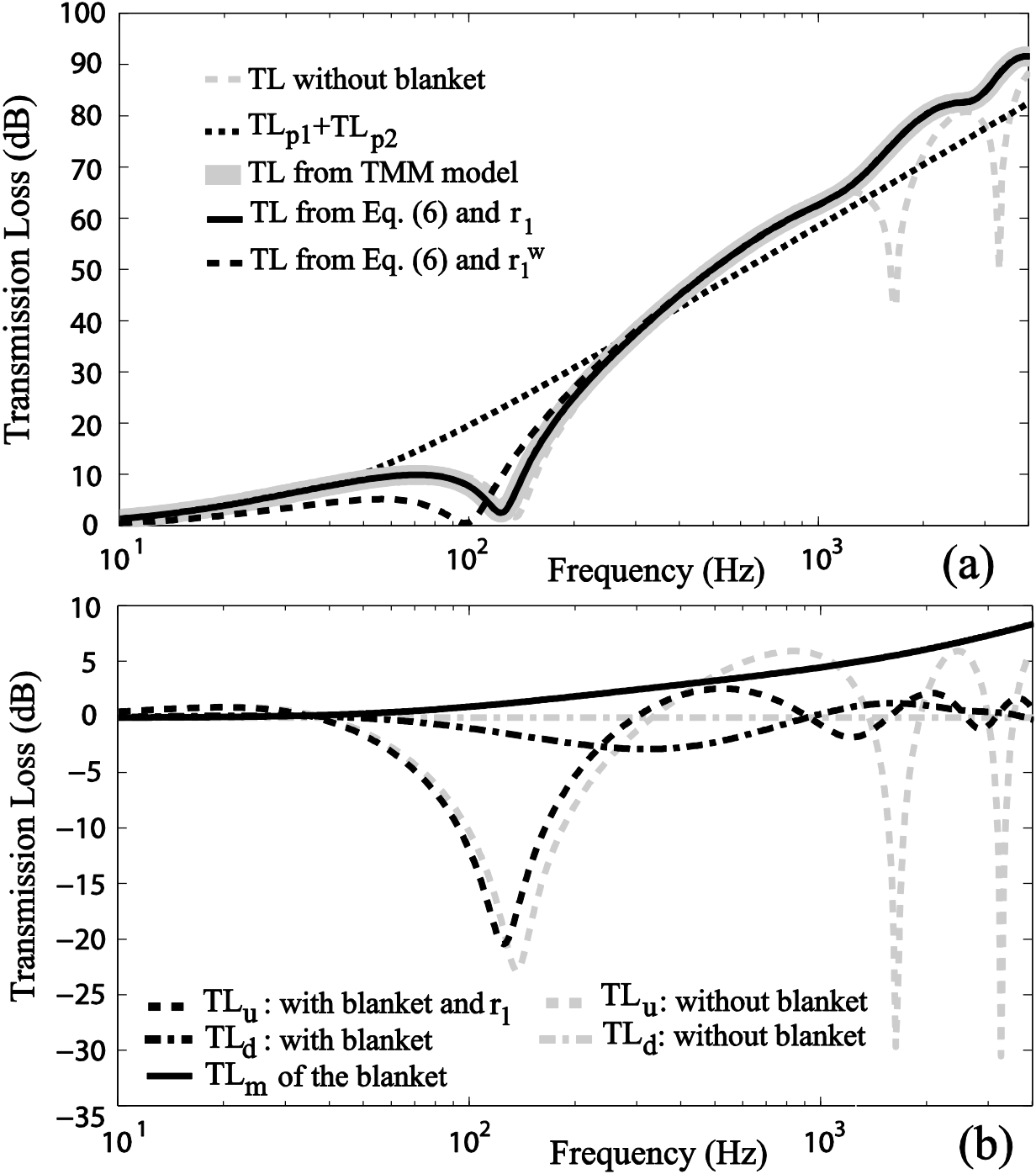

*Figure 3 : (a) Normal incidence sound transmission loss of the empty structure or filled in with the multilayer blanket "MaterialA/Screen/MaterialC"; (b) sound transmission loss contributions of the blanket or of the empty cavity.*

Regarding the sound transmission loss of the double panel structure with and without multilayer inside the air-gap (see Fig. 3(a)), it is found as expected that the multilayer blanket mainly attenuates the dips of insulation controlled by the cavity resonances around 1.6 kHz and 3.3 kHz and improves the insulation at high frequencies. At low frequencies, the blanket reduces the dynamic stiffness of the air between the plates and then slightly decreases the "mass/spring/mass" resonance frequency.

According to the proposed expression of $TL$ (Eq. (6)), it is now possible to investigate the various contributions $TL_u$, $TL_d$ and $TL_m$ (see Fig. 3(b)). When the double panel structure is empty, it is found that the contribution of the multiple reflections in the downstream cavity, $TL_d$, is null and on the contrary, the contribution of the multiple reflections in the upstream cavity $TL_u$ accounts for all acoustical and mechanical effects: acoustic resonances and "mass/spring/mass" resonance. When the multilayer absorbing blanket is present in the structure, the main effects are thus visible on the $TL_u$ contribution for which the insulation dip at the cavity resonance is greatly reduced and the "mass/spring/mass" resonance slightly shifted toward lower frequency. The fact that $TL_u$ accounts for the various couplings between the two panels and not $TL_d$ is due to their respective expression (see Eqs. (7),(8)) in which $TL_u$ accounts for the presence of panels 1 and 2 by $r_{p1}$ and $r_1$ respectively, and $TL_d$ only accounts for the presence of one panel by $r_{p2}$. Indeed, $r_2$ is calculated when the sound absorbing blanket is backed by a semi-infinite air layer and is close to 0 as shown in Fig. 4 (see column 2). This confirms that the reflection coefficient of the rear surface of the material $r_2$, is different from the one of the front surface $r_1$, as mentioned by Salissou and Panneton in reference[20].

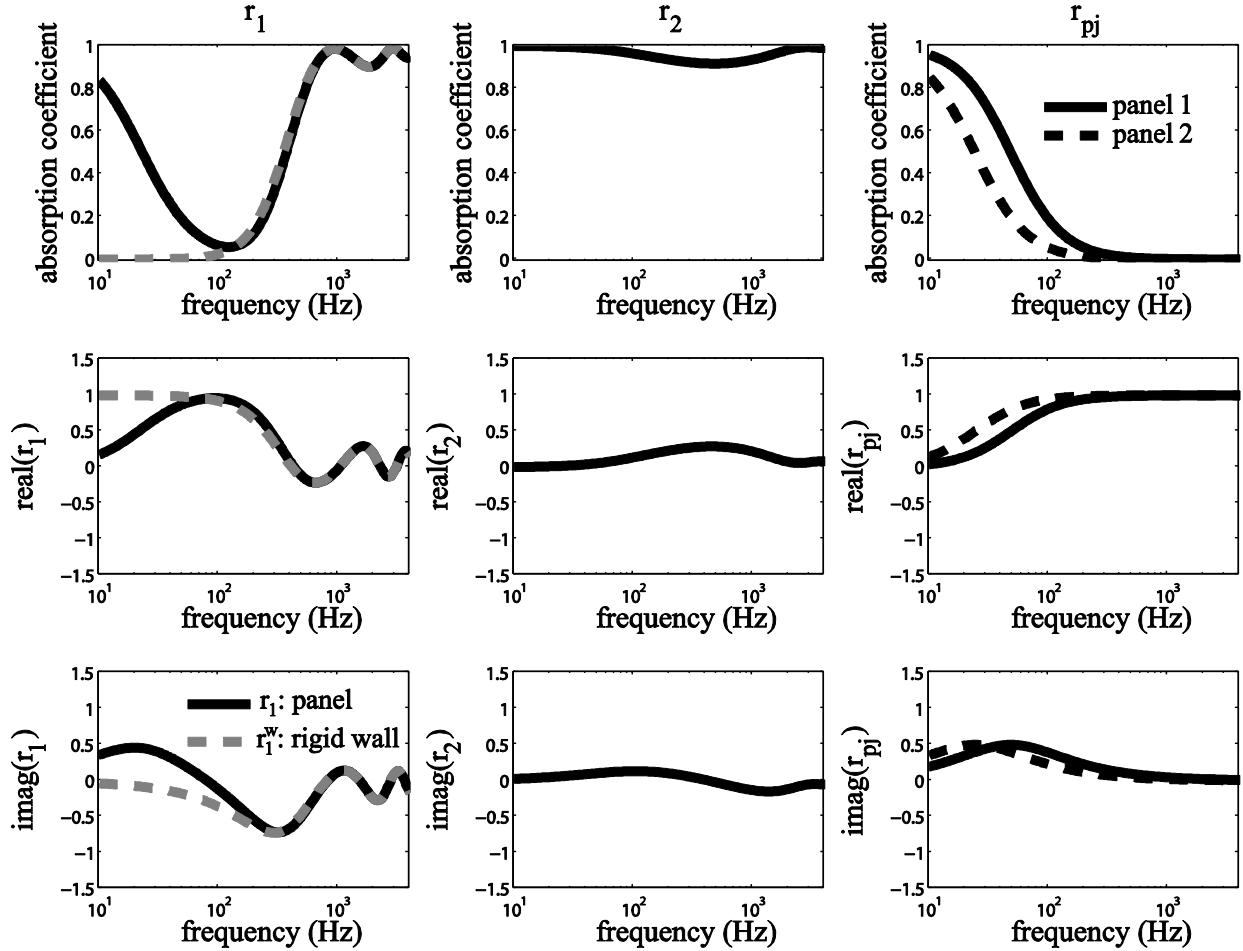

*Figure 4 : Simulation of the reflection coefficients for the multilayer "MaterialA/Screen/MaterialC".*

Fig. 3(b) shows that, in the case of this multilayer blanket, the contributions of $TL_u$ and $TL_d$ are of the same amplitude above 200 Hz. It is also shown that these two contributions are inferior to the contribution of the transmission loss of the multilayer blanket $TL_m$. Whereas the sound transmission loss of the multilayer increases continually with frequency, the sound transmission losses due to upstream and downstream absorptions present slight oscillations around 0dB. Finally, comparing now all the sound transmission loss contributions of Eq. (6) and presented in Figs. 3(a) and 3(b), it is found as expected that the contribution of the two aluminum panels is predominant due to mass effect. Still, the sound package allows a 5-10 dB improvement in the high frequency range.

It is worth mentioning that even if the previous simulations have been carried out with 25mm and 30mm upstream and downstream cavities respectively, the presented model is also valid for small cavities, for example $D_1=D_2=$1mm. On a practical note, the proposed expression of the global *TL* (Eq. (6)) can also be used to estimate the normal incidence sound transmission loss of a double panel structure filled with a multilayer sound absorbing blanket from the measurements of the surface density of the two panels and the measurements of the transmission and reflection coefficients of the multilayer blanket $TL_m$, $r_1^w$ and $r_2$ [18-21]. In this case, the measurement of all the non-acoustic properties (i.e., porosity, static airflow resistivity, tortuosity, viscous and thermal characteristic lengths) of each component of the multilayer filling the double panel structure and required generally in the models is not necessary. However, it is important that the experimenter ensures that the multilayer blanket behaves as an equivalent fluid (rigid or limp) inside the impedance tube with no contribution of the frame elastic behavior and no leakage effects[22-25].

**B. Comparison of the acoustic behaviour for three different monolayer porous materials**

The influence of the acoustic behaviour of the absorbing blanket, i.e. contributions of the sound absorption inside the double panel structure and sound transmission loss of the blanket, is now investigated more in details. Here, three 50mm-thick monolayer porous materials with different pore geometries are used as blanket inside the double panel structure. Material A and B are low and high static airflow resistivity plastic foams, respectively, both with a stiff and low density skeleton. Material C is low density fibrous material with a soft skeleton and a low static airflow resistivity. These three materials are frequently used in aerospace and building applications for thermal and sound insulation. Properties of the materials are listed in Table I. These materials have been selected because of their distinct acoustic behaviour related to the porous microstructure, i.e. two materials are foams constituted of a continuous arrangement of cells (first is reticulated, the second is not), and the other material is fibrous constituted of a discontinuous stack of fibres. Finally, despite of their different microstructure, material A shows (i) similar sound transmission loss behaviour to material B which allows us to focus on the absorption contributions ($TL_u$, $TL_d$) and (ii) similar sound absorption behaviour to material C which allows us to focus on the sound transmission loss contribution ($TL_m$). In order to emphasize the effects of the porous layer, the double panel structure is considered symmetric here, i.e. $D_1 = D_2 = 25$mm and $m_{s1} = m_{s2} = 2.742$ kg/m$^2$. Since materials A, B and C are monolayer blankets, the reflection coefficient $r_2$ is derived this time using Eq. (14). Furthermore, note that $TL_u$ is derived using the reflection coefficient $r_1$ and not $r_1^w$ since the effects of the rigid wall assumption has already been investigated in the previous section.

Fig. 5 (a) shows the normal incidence transmission loss of the double panel structure with either foam A or foam B and the normal incidence sound transmission loss of the empty structure. Fig. 5(b) shows the contributions of the sound transmission loss of the porous material A and B ($TL_m$) and their upstream ($TL_u$) and downstream ($TL_d$) absorption contributions. It is shown that, even if the two materials have similar sound transmission loss behaviour (see solid lines Fig. 5(b)), the absorption contributions can be greatly different (see dashed and dash-dotted lines Fig. 5(b)) and thus lead to a different sound transmission loss performance of the double panel structure (see Fig. 5(a)). In this case, the lack of performance of the sound absorption behaviour of material B compared to material A in the medium frequency range, i.e. between 400 Hz and 1200 Hz (see dashed and dash-dotted grey curves Fig. 5(b)), leads to a lack of performance of the sound transmission loss of the double panel structure when material B is present (see Fig. 5(a)). In this frequency range, the sound transmission loss of the empty structure is even better: the transmission loss with foam B is 8 dB lower at 900Hz.

To illustrate the absorption behaviour difference between material A and B, figure 6 presents the normal incidence sound absorption associated to the reflection coefficient $r_1^w$ for the three materials backed by the air cavity $D_2$ and a rigid and immobile wall. These simulations are derived from Eq. (13). It is shown that compared to material A and C, material B presents a gap in performance in the mid frequency range. It is worth commenting that the frequency band of this sound absorption dip is not exactly the one observable in the normal sound transmission loss of the double panel structure (Fig. 5(a)). Indeed, according to Eq. (7), the transmission coefficient $TL_u$ related to the multiple reflections effect inside the upstream cavity does not only depend on the absorption behaviour of the "porous/downstream cavity/panel 2" sub-system (related to $r_1$) but also on the size of the upstream cavity $D_1$.

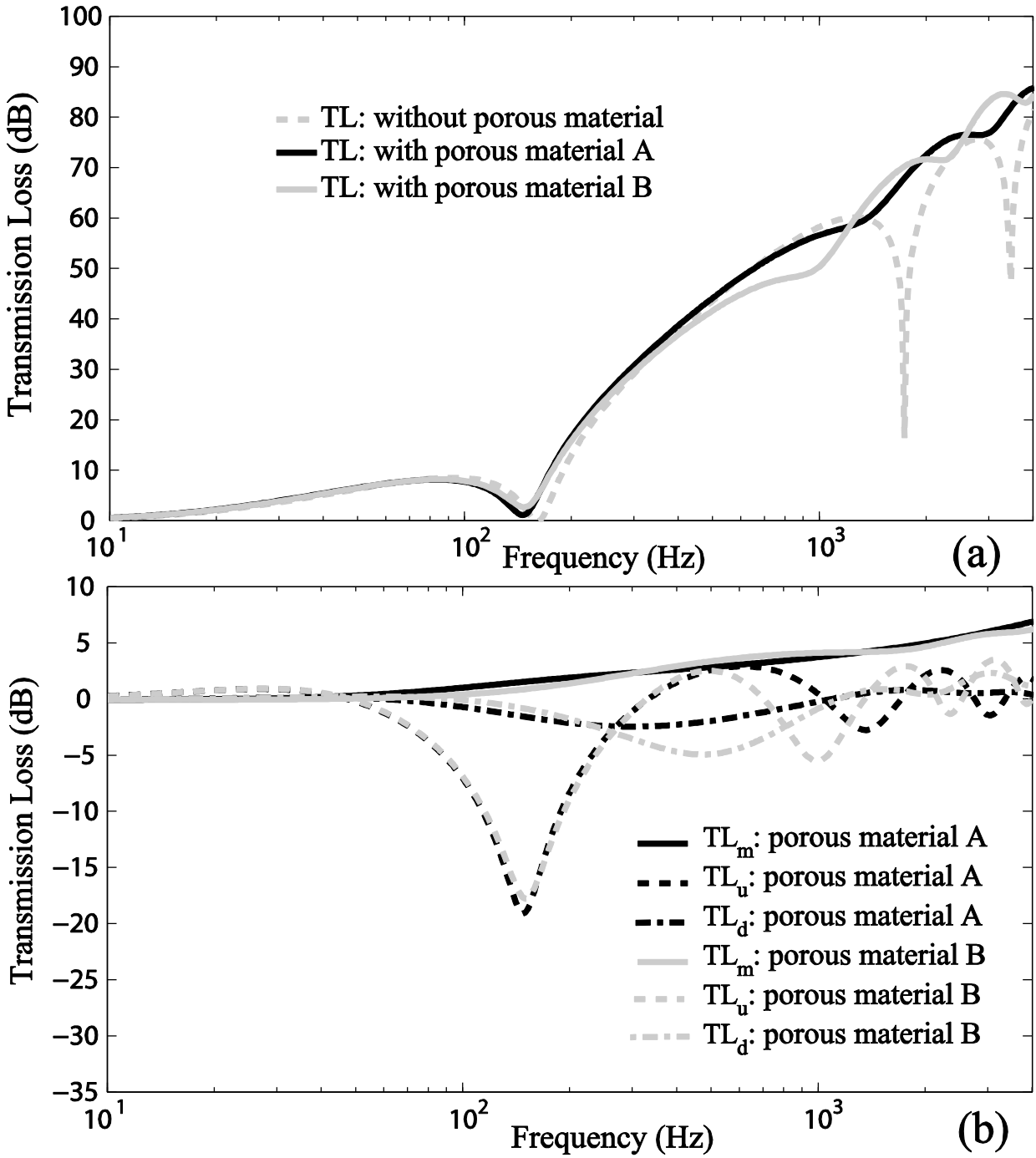

*Figure 5 : (a) Normal incidence sound transmission loss of the double panel structure; without porous material, with material A, with material B; (b) sound transmission loss contributions: with material A, with material B.*

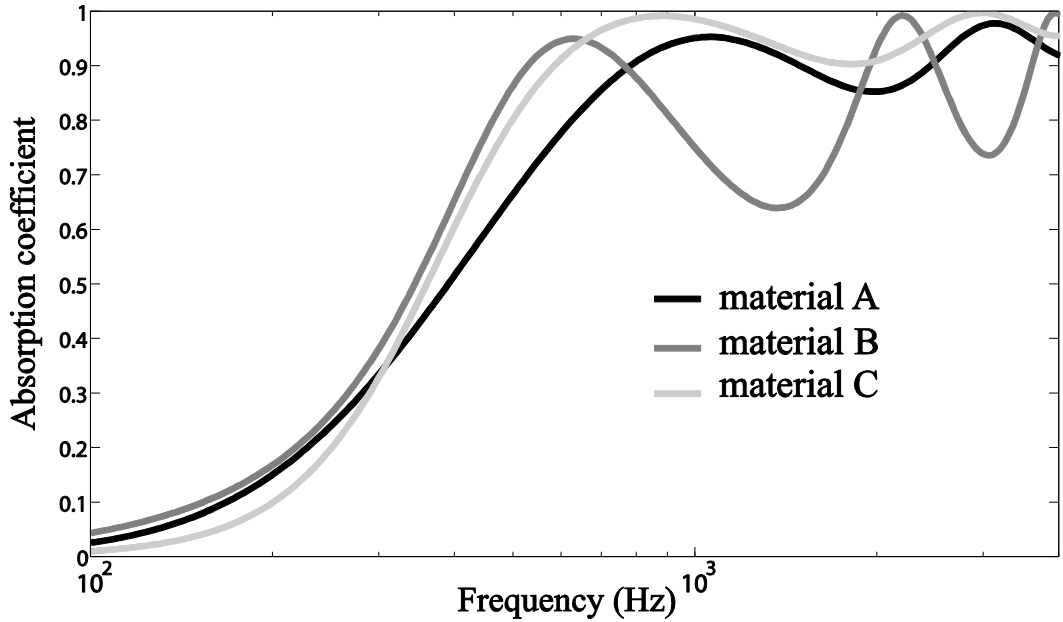

*Figure 6 : Normal incidence sound absorption coefficient of a 50mm thick porous material backed by a 25 mm thick air cavity $D_2$ and a rigid and immobile wall ($\alpha = 1 - |r_1^w|^2$).*

Fig. 7 (a) shows the normal incidence sound transmission loss of the double panel structure including foam A or fibrous C and the one of the empty structure. Fig. 7(b) shows the contributions of the sound transmission loss of the porous material A and C ($TL_m$) and their upstream ($TL_u$) and downstream ($TL_d$) absorption contributions. It is shown that, even if the two materials have similar sound absorption behaviour inside the double panel structure (see dashed lines Fig. 7(b)), the sound transmission loss of material C is slightly greater than the one of material A (see solid lines Fig. 7(b)), which leads to a greater sound transmission loss performance of the double panel structure when material C is present (see Fig. 7(a)). In this case, the higher transmission loss performance ($TL_m$) of material C compared to material A cannot be attributed to a mass effect as its bulk density is lower, but rather to an improved visco-thermal dissipation mechanism due to pore geometry.

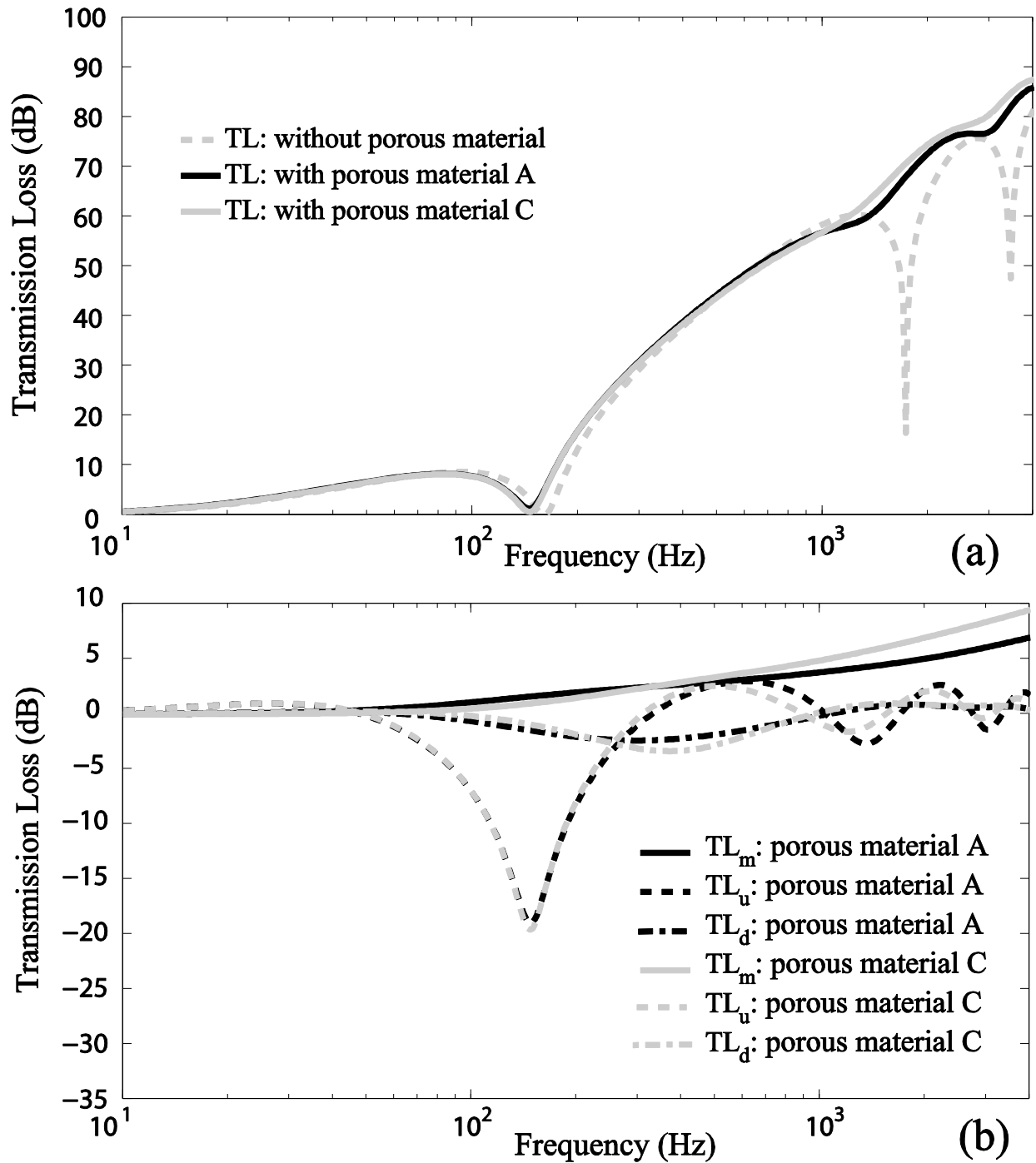

*Figure 7 : (a) Normal incidence sound transmission loss of the double panel structure; without porous material, with material A, with material C; (b) sound transmission loss contributions: with material A, with material C.*

# IV. CONCLUSION

In this paper, the different acoustic contributions of a sound absorbing blanket placed between two thin panels to the normal incidence sound transmission loss of the double panel system has been investigated. For this purpose, a simple analytic expression of the normal incidence sound transmission loss of the double panel structure is proposed in terms of three main contributions (see Eq. (6)): sound transmission loss of the panels, sound transmission loss of the blanket and sound absorption due to multiple reflections inside the structure. The acoustic contributions of the sound absorbing blanket are then investigated in the case of four different blankets frequently used in the context of transport applications: a non-symmetric multilayer made of a screen in sandwich between two porous layers and three symmetric porous layers having different pore geometries. It is shown that; (i) at high frequencies, the transmission loss contribution of the blanket is preponderant compared to the absorption contributions; (ii) at the cavity resonance frequencies, the absorption contribution allows to attenuate the dips of insulation, (iii) at medium and low frequencies,  for  porous layers showing poor absorption performance, the absorption contributions in the air-gaps can decrease the sound transmission loss performance of the double panel structure which can be even better in the case of an empty structure. Using the proposed expression of Eq. (6), one can accurately estimate the normal incidence sound transmission loss of a double panel structure by measuring few acoustic properties of the absorbing blanket with classical impedance tube techniques. This could be used as an alternative to the classical models which require the measurement of all the non-acoustic properties (e.g. Biot properties) of each layer constituting the blanket. The proposed expression also represents a practical way to optimize the TL of such structures by concentrating on the sound package.

## ACKNOWLEDGEMENTS

The authors would like to thank the National Sciences and Engineering Research Council of Canada (NSERC) for providing financial support.